\documentclass[11pt, a4paper]{article}

\usepackage[english]{babel}
\usepackage[utf8]{inputenc}
\usepackage[T1]{fontenc}

\usepackage[a4paper,top=3cm,bottom=2cm,left=3cm,right=3cm,marginparwidth=1.75cm]{geometry}

\usepackage{amsmath}
\usepackage{graphicx}
\usepackage[colorinlistoftodos]{todonotes}
\usepackage[colorlinks=true, allcolors=blue]{hyperref}
\usepackage[square,sort,comma,numbers]{natbib}
\usepackage{amssymb}
\usepackage{float}
\usepackage{caption}
\usepackage{appendix}
\usepackage{dcolumn}
\usepackage{booktabs}
\title{Labor Market Outcomes and Early Schooling: Evidence from School Entry Policies Using Exact Date of Birth\footnote{We thank our colleagues and teachers Cecília Machado, Romero Rocha, Rudi Rocha, Naércio Menezes-Filho and Bruno Oliva for comments and suggestions. All errors and omissions are the sole responsibility of the authors.}}

\author{Daniel Duque\thanks{Institute of Economics, Federal University of Rio de Janeiro, Av. Pasteur, 250 - Urca, Rio de Janeiro - RJ, Brazil, 22290-902. E-mail: daniel.duque@ppge.ie.ufrj.br. Declarations of interest: none.} \and Pedro Cavalcante \thanks{Department of Economics, Federal Fluminense University,  Rua Alexandre Moura, Bloco F, 8 - São Domingos, Niterói - RJ, Brazil, 24210-200. Corresponding author, e-mail: pc\_oliveira@id.uff.br. Declarations of interest: none. }}

\begin{document}
\maketitle
\begin{abstract}

We use a rich, census-like Brazilian dataset containing information on spatial mobility, schooling, and income in which we can link children to parents to assess the impact of early education on several labor market outcomes. Brazilian public primary schools admit children up to one year younger than the national minimum age to enter school if their birthday is before an arbitrary threshold, causing an exogenous variation in schooling at adulthood. Using a Regression Discontinuity Design, we estimate one additional year of schooling increases labor income in 25.8\% - almost twice as large as estimated using mincerian models. Around this cutoff there is also a gap of 9.6\% on the probability of holding a college degree in adulthood, with which we estimate the college premium and find a 201\% increase in labor income. We test the robustness of our estimates using placebo variables, alternative model specifications and McCrary Density Tests.  

\end{abstract}

\textbf{JEL Classification}: I24 I26 I28 J24 J31

\textbf{Keywords}: Education, College Premium, RD Design

\clearpage

\section{Introduction}\label{introdução}

Brazil, despite improvements in the last 20 years, remains a poorly educated country, even compared to its neighbors. Among the largest countries in Latin America, only Colombia has educational levels similar to Brazilians. The gap between its average years of schooling and leaders of the region is around 25\%.

Although the strong role of human capital in wage formation - not only in Brazil - is consensual (\cite{langoni1973distribuiccao}, \cite{barbosa2010evoluccao}, \cite{de2014evoluccao}), there is little empirical evidence on the country's labor returns of education.  \cite{griffin1993rates} estimates from a mincerian equation the rate of return to an additional year of schooling between 12.8 and 15.1\%. \cite{stefani2006returns} estimated a 28\% increase in wages with data from 1996. However, there is a general lack of estimates using evidence from natural experiments or with emphasis on research design, with a credible causal identification strategy, as described in \cite{angrist2010credibility}. Estimating returns to schooling returns and college premiums in lower middle-income countries such as Brazil is an immediately policy-relevant question. 

In this paper, we present causal evidence of educational returns on labor market using a Brazilian usual school entry rule as an instrument for education in adulthood. We exploit the fact that the year in which a person starts school is a non-monotonic function of date of birth. Our results show that the school entry policy has significant effects on schooling in adulthood. Brazilians aged between 16 and 34 who used to live in state capitals or urban areas when they were 15 who were born just before the most usual school entry date (March 31) on average have close to one additional year of study, in comparison to those born just after the threshold. For those occupied in the labor market, the expected difference is of 0.77 years of schooling.

Getting enrolled in primary school earlier also has an impact on the probability of holding a college degree in adulthood. In the first sample mentioned above, those born just before the threshold were around 9\% more likely to hold a college degree than the ones born just after the school entry date and those on the labor force who used to live in urban areas at the age of 15 are 9.6\% more likely. Using these discontinuities as instruments for educational level, we estimate the impact of an additional year of schooling and college attendance on labor market outcomes. Our findings show that being marginally more schooled has an effect of 25.8\% over wages, but none over labor supply and probability of being employed. Having tertiary education has a labor income return of 201\%, but also shows no effects on employment. 

Because of law Nº 9.394, of December 20 (1996) of the Brazilian Federal Constitution, children must attend school from the age of 6, however, since school enrollment date is in the first or second month of the year, schools accept students under the obligatory age if their birthday is before a given threshold. In 2010, the Brazilian National Education Council established March 31 as a national policy for school entry rule of children younger than the minimum age for primary school at the time of enrollment. Prior to that, each state or municipal education system decided its date of birth threshold to ingress in the year's class when children were under the age of primary school enrollment. However, surveys conducted by \cite{todospelaeduc2015} showed that most of major cities in Brazil were adopting March 31 as threshold. 

This paper uses data from the Brazilian National Sample Household Survey (PNAD), a yearly survey conducted by the Brazilian Institute of Geography and Statistics, which was discontinued in 2015. Unlike many other surveys with schooling-related variables, PNAD contains exact date of birth data. More specifically, the National Sample Household Survey has a supplement focusing on Socio-Occupational Mobility in its 2014 edition, in which surveyees were asked in which state they used to live at the age of 15 and if it was a state capital, or an urban area. We use this information to restrict our sample to cities where we expect there's better quality enforcement of school entry rules.

Our paper contributes to the literature showing that, despite no apparent effect of being born just before the most usual threshold in Brazil, March 31, even for the population between 16 and 34 currently living in urban areas, that is likely due to a common rural-urban migration in adulthood. Using data from PNAD's 2014 edition supplement we show that the discontinuity around the threshold is statistically significant for a sample comprised only of people who used to live in state capitals and/or urban areas when 15 years old, which fits our intuition that urban areas are more likely enforce such school entry date rules.  

\cite{10.1257/aer.101.1.158} states that using school entry rules as a source of exogenous variation of schooling in adulthood has its limitations. That happens because not all children enter school in the year predicted by primary school entry policy - the parents of the ones born before the threshold may alternatively keep them at home or preschool for another year, as it is also possible for the ones parenting children born after the school entry date to make a petition for their child to start school a year before typically allowed, or even put their child in a private school. This means that it's possible for our estimates to be disproportionate relative to low-income families, whose parents are more likely to comply with school entry rules. 

Our paper has an additional limitation: the lack of precise primary school entry policies at a municipal level. Our estimates are confined only to cities which were establishing March 31 as their threshold for enrolling a child before he or she was 6 years old. It's also relevant to clarify that school entry rules naturally can only affect education in adulthood for those who are still in school or school dropouts, having then fewer years of education if he or she were first enrolled late \citep{angrist1992effect}.

This paper is organized as follows. In Section ~\ref{teoria} there's a brief discussion on the human capital theory and the alternative ways in which education impacts labor market outcomes, discussing the literature on this issue, specially for Brazil. In Section~\ref{metodologia}, we present our identification strategy, followed by a description in Section~\ref{dados} of the data we used. Results are presented in Section~\ref{resultados} and robustness checks in Section~\ref{robustez}. Section~\ref{conclusão} concludes.

\section{Education and Labor Markets}\label{teoria}

There are three primary channels by which education can impact labor income. The first one is by raising productivity via human capital accumulation (see \cite{schultz1961investment} and \cite{becker1994human}). Not only does human capital lower cost of information acquisition \citep{rosenzweig1995there}, but also as workers acquire skills, experience and knowledge, a wider array of what they can produce, and how efficiently, becomes available. \cite{doi:10.1111/0034-6527.00325} and \cite{RePEc:aea:aecrev:v:84:y:1994:i:5:p:1157-73} provide some empirical evidence and rates of return of education.

The second channel is through peer effects, both at firm and sector level. As more educated workers join a particular firm or sector, productivity for other workers rises \citep{doi:10.1086/497818}. This can happen either due to direct learning from other workers or due to some sort of spillover effect, e.g. a recently-hired, more educated worker implementing cost-saving measures.

The third is through signaling \citep{spence1978job}. The idea is that, due to asymmetric information, high school diplomas, college degrees and other forms of education certificates act as signals. As they're costly to acquire, employers are more willing to hire workers with more, better signals and also to do so for higher wages. \cite{hamalainen2008signalling} and \cite{castagnetti2005educational} provide some empirical evidence for this phenomenon.

Overall, education has a very well documented, positive effects on wages \citep{dickson2011economic} and positive externalities. \cite{doi:10.1093/qje/qjv004} documents increases in political participation and standards of living not only for more educated people but local spillovers of the benefits (see \cite{moretti2004human} for a comprehensive review of this topic).

\section{Methodology}\label{metodologia}

Our empirical strategy relies on "Fuzzy" Regression Discontinuity Design estimates. On that matter, first, we'll model schooling as a non-monotonic function of date of birth. Let $S_i$ be schooling (measured by years of formal education), $N_i$ be date of birth, $\alpha(N_i)$ be a continuous function $\forall N_i \neq = 0 $, implying a cutoff at $0$. To be more precise, $\nexists \lim_{{N_i}\to 0} \alpha(N_i)$. Also, let $D_i=1$  if $N_i > 0$ and $\mu_i$ be a normally distributed error term. Parameter $\beta$ measures the local average treatment effect (LATE) of the primary school entry rule. Consider the model:

\begin{equation}
S_i = \alpha(N_i) + \beta_1 D_i + \mu_i
\end{equation}

In a Fuzzy RDD, first stage estimation is one in which schooling is partially affected by the discontinuity. The second stage follows this model:

\begin{equation}
Y_{ic} = \beta_0 + \beta_2 \hat{S}_{ic} + \epsilon_i
\end{equation}

Where $Y_{ic}$ is the outcome of interest. In this paper, we will assess the impact of education over employment and labor income. We estimate our model with the Fuzzy RD design proposed by \cite{calonico2014robusteconometrica}, with robust bias-corrected confidence intervals. We use schooling as an instrument and date of birth as running variable, with the primary school entry policy for children under aged when getting enrolled as cutoff. 

RD Design is most similar to a randomized experiment, so that it’s possible to argue that the coefficient found in the regression does represent a causal effect. Despite the fact its LATE estimates may be interpreted as only applicable to the subpopulation of individuals at the discontinuity threshold, and uninformative about the effect anywhere else, \cite{lee2010regression} argue that, in the presence of heterogeneous treatment effects, RD estimates can be interpreted as a weighted average treatment effect, with weights relative to the ex-ante probability that the value of an individual’s assignment variable (in this case, date of birth) will be around the threshold.

In RD Design, the choice of the threshold for the cutoff is quite determinant for robustness of the estimates. So first we look at the legislation on the topic. Table ~\ref{datas_corte_estaduais} lists 12 states and two cities by legal primary school entry dates. Other states and many municipalities use different - many informal - rules to decide local entry dates. As observed in Table~\ref{datas_corte_estaduais}, at least 9 of the 27 brazilian States have legislation establishing the threshold of the primary school entry date at March 31, besides Brazil's biggest city, São Paulo. April 30 and June 30 also appear as other legislated dates for letting a child be enrolled before he or she is 6 years old. 

\begin{table}[H]
\caption{Thresholds by state. Source: State legislations}
  $  \begin{array}{c c c } 
 \hline
 \text{March 31} & \text{April 30} & \text{June 30}  \\ [0.5ex] 
 \hline\hline
 \text{Distrito Federal} & \text{Mato Grosso} & \text{Ceará}  \\ [0.5ex]
 \hline
 \text{São Paulo (state capital)} & \text{São Paulo (Atibaia city)} & \text{Minas Gerais}  \\ [0.5ex]
 \hline
 \text{Rondônia} &  & \text{São Paulo (rest of the state)} \\ [0.5ex]
 \hline
 \text{Tocantins} &  &  \\ [0.5ex]
 \hline
 \text{Rio Grande do Sul} &  &  \\ [0.5ex] 
 \hline
 \text{Santa Catarina} &  &  \\ [0.5ex] 
 \hline
\text{Goiás} &  &  \\ [0.5ex] 
 \hline
\text{Pernambuco} &  &  \\ [0.5ex] 
 \hline
\text{Pará} &  &  \\ [0.5ex] 
 \hline
 \end{array}$
\label{datas_corte_estaduais}
\\[1.5pt] 
\end{table}

\begin{figure}[H]
\includegraphics[width=\textwidth] {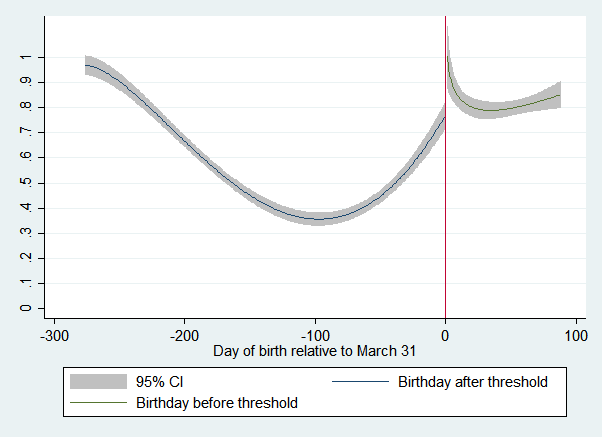}
\caption{Share of 6 years old children attending primary school}
\label{crianças de 6 anos na escola}
\end{figure}

In figure~\ref{crianças de 6 anos na escola}, we  show a polynomial function of the probability of 6 years old children born two months before and after March 31 being enrolled in primary school by September. Dates of birth were numerically modified so that for those born at the threshold, it gets a zero value, with negative values for posterior dates and positive values for dates preceding it. Figure~\ref{crianças de 6 anos na escola} shows that there's a high discontinuity on the share of 6 years old children enrolled in Primary School/ 1st grade around March 31, as expected. That means at least a large number of urban cities do enforce this school entry date as a threshold.    

\begin{figure}[H]
\includegraphics[width=\textwidth] {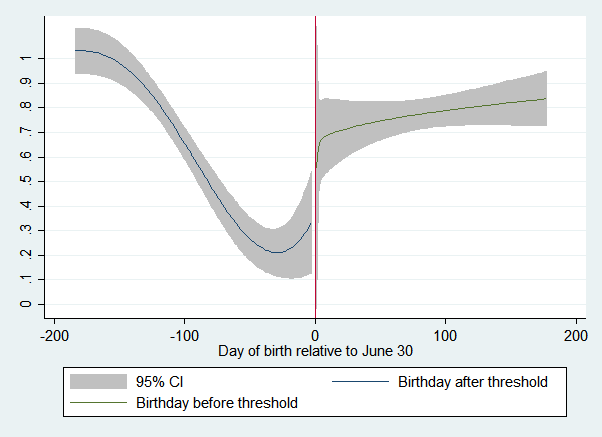}
\caption{Share of 6 years old children attending primary school in Minas Gerais}
\label{crianças de 6 anos na escola MG}
\end{figure}

\begin{figure}[H]
\includegraphics[width=\textwidth] {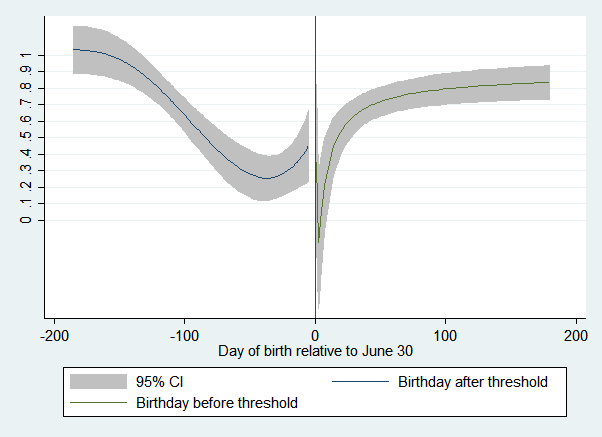}
\caption{Share of 6 years old children attending primary school in Ceará}
\label{crianças de 6 anos na escola CE}
\end{figure}

\begin{figure}[H]
\includegraphics[width=\textwidth] {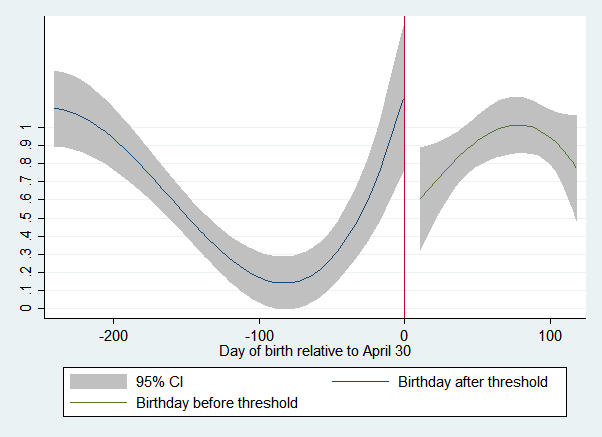}
\caption{Share of 6 years old children attending primary school in Mato Grosso}
\label{crianças de 6 anos na escola MT}
\end{figure}

For other states with legislation on primary school entry date, figures~\ref{crianças de 6 anos na escola MG},~\ref{crianças de 6 anos na escola CE} and~\ref{crianças de 6 anos na escola MT} suggest that there's a discontinuity in school attendance for children born around June 31 in urban cities of Minas Gerais, but not for the ones in Ceará. Mato Grosso has also not shown any statistically significant discontinuity around April 30. These two last cases may be due to lack of enforcement of the written legislation on primary school entry date, what Brazilians usually refer to as a "law that doesn't stick".\footnote{This is somewhat characteristic of weakly institutionalized societies and likely a product of Brazil's culture of evaluating congressmen by the number of laws they have approved, which creates an incentive to propose several laws that may or may not be actually enforced. See \cite{pires2009estilos} for a more detailed discussion.} 

\cite{pena2014impact} utilized Brazilian School Census data to find discontinuities on enrollments by date of birth to determine each city's heuristic on the topic. However, PNAD does not have municipality-level information, constraining our precision on the threshold choice. According to a survey by \cite{todospelaeduc2015}, at least 19 of the 27 Brazilian state capitals were adopting the March 31 date as primary school entry policy, as were other 718 municipalities from the 1230 that answered to the survey. With most of population living in cities with the same policy, we chose to use March 31 as the our threshold for all states, even the ones with specific legislation on this topic. We provide some placebo tests with different dates (even fake ones) for robustness.

\section{Data and Sample}\label{dados}

We've used data from the Brazilian National Sample Household Survey (PNAD), a cross-section of nearly 350,000 individuals conducted yearly from the mid 70s to 2015 by the Brazilian Institute of Geography and Statistics (IBGE). PNAD has data on numerous characteristics of the population, such as labor conditions, income and education. As mentioned before, PNAD is not stratified to have representative information at the municipality level, only on the State and Metropolitan Region levels. Thereby, it's not possible to identify the city each interviewee was living in. However, it does identify whether the city was in an urban or rural area.

Almost every year, PNAD has additional questions regarding a specific topic, called a Thematic Supplement. In 2014, it had additional data about socio-occupational characteristics of the population with at least 16 years of age and their families, but when they were 15 years old. This way, we have data informing whether they lived in one of the 26 state capitals at that age, the state they were living in, and whether the city was in a urban or rural area. The sample with this additional data has almost 56,000 individuals. 

In Brazil, only at the end of the twentieth century, with a delay of almost a century in relation to developed countries, compulsory primary education was virtually universalized with regard to access \citep{portela2007universalizaccao}. Due to this limitation, we will restrict our sample to population born after 1980, possibly getting integrated in the education system in the second half of the 80's. This restriction reduces our sample to about 21,200 individuals. Our last restriction is due to high heterogeneity of public school rules and lack of school entry policy enforcement in rural areas. In order to control for it, our sample will be restricted to population living in urban areas in 2014 or lived in such areas when they were 15 years old. This seems reasonable, as more urbanized areas should have better enforcement.

\section{Results}\label{resultados}

\begin{figure}[H]
\includegraphics[width=\textwidth] {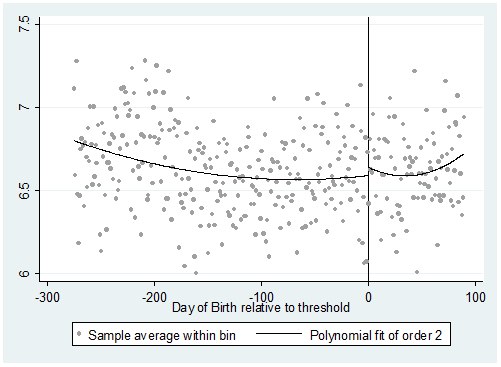}
\caption{Years of schooling and date of birth for people between 16 and 34 currently living in urban areas.}
\label{escolaridade_2_descontinuidade URBANO}
\end{figure}

Figure ~\ref{escolaridade_2_descontinuidade URBANO} summarizes our results. When we consider all population between 16 and 34 currently living at urban areas in 2014, there's no discontinuity in years of study around the threshold, illustrated by figure~\ref{escolaridade_2_descontinuidade URBANO}. This may happen because part of the population who had grown in rural areas migrated to urban areas when older, looking for labor market opportunities. This way, our sample partially includes those who lived in regions with high heterogeneity of public rules and lack of enforcement, adding variance to our estimates. 

\begin{figure}[H]
\includegraphics[width=\textwidth] {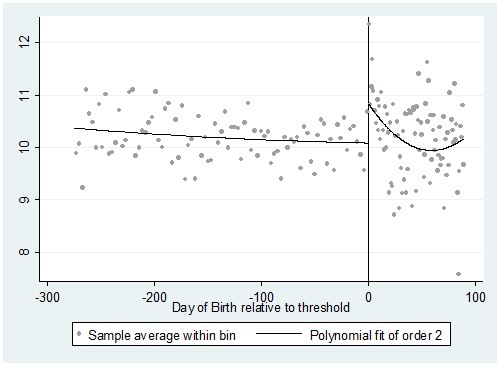}
\caption{Years of schooling and date of birth for people between 16 and 34 who lived in a state capital when aged 15.}
\label{escolaridade_2_descontinuidade capital 15}
\end{figure}

\begin{figure}[H]
\includegraphics[width=\textwidth] {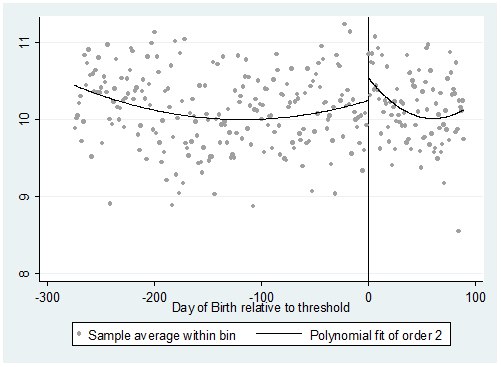}
\caption{Years of schooling and date of birth for people between 16 and 34 who lived in a urban city when aged 15.}
\label{escolaridade_2_descontinuidade urbano 15}
\end{figure}

However, when we use the information in the 2014 edition's supplement to restrict our sample to the population between 16 and 34 years old who used to live in a state capital when 15, we do find a high and significant discontinuity in years of study around the cutoff, as showed in figure~\ref{escolaridade_2_descontinuidade capital 15}. Besides, using a larger sample, with people between 16 and 35 years old who used to live in an urban area when 15, figure~\ref{escolaridade_2_descontinuidade urbano 15} shows that we still find a significant - but smaller - discontinuity on years of schooling. 

Then, we run a Fuzzy Regression Discontinuity Design to check the effect of years of study on labor market variables. We use years of study instrumented by its discontinuity around the cutoff in the running variable as independent variable to explain labor income (for occupied population), hours in labor (for all sample) and probability of being occupied (for economically active population). Results are summarized in table~\ref{results}.

In the first stage, we find a 10\% significant expected increase of 1.12 years of study for the sample of occupied population between 16 and 34 years old who used to live in state capitals when 15, and a slightly smaller effect of 1.04 (still significant at 10\%) for all sample and no significant effect for the economically active population. In the second stage, none of estimates were statistically significant. This result may be due to our limitation in the number of observations in our sample. When using a larger sample, including anyone between 16 and 34 that lived in urban areas when 15 years old, we do find an expected increase of 0.77 years of study for the occupied population born just before March 31, statistically significant at 5\%. For all sample, our estimate goes to 0.65, also reducing its significance to 10\%, and losing it in the sample only with economically active population. 

In the second stage for the occupied population, we find that an exogenous increase of one year of study is expected to raise labor income in 25.8\%, but without any effect on hours of work and probability of being employed. These last results, however, might be due to the insufficient significance from the first stage. We also check whether being born just before March 31 has an impact on the probability of holding a college degree in adulthood. In our samples with 16 and 34 years old population currently living in urban areas or that lived in a state capital when 15 showed no significant effect of being born just before March 31 on years of study - this second result, again, possibly being due to lack of mass in the sample.  

\begin{table}[H]
\caption{Fuzzy RDD estimates.}
  $  \begin{array}{  l  l  l  l  l  }
   \hline
	 Sample & Stage & ln(wages) & ln(hours) & p(emp) \\ \hline\hline
\text{Currently living in urban areas}  & 1st & .280 & .272 & .205 \\ 
	outcome: \text{Years of Study} & 2nd & -.057 & -.047 & -.192 \\ \addlinespace
	\text{Lived in a state capital when aged 15} & 1st & 1.117^{*} & 1.04^{*} & .908 \\ 
	outcome: \text{Years of Study} & 2nd & .083 & .103 &  -.157 \\ \addlinespace
	\text{Lived in urban area when aged 15} & 1st & .773^{**} & .645^{*} & .528 \\ 
	outcome: \text{Years of Study} & 2nd & .258^{**} & .181 & -.139 \\  \addlinespace
	\text{Currently living in urban areas}  & 1st & .021 & .014 & .017 \\ 
outcome: \text{Probability of holding college degree}	 & 2nd & -.785 & -.958 & -2.342 \\ \addlinespace
	\text{Lived in a state capital when aged 15} & 1st & .054 & .048 & .043 \\ 
outcome: \text{Probability of holding college degree}	 & 2nd & 1.726 & 2.177 & -1.254 \\ \addlinespace
	\text{Lived in urban area when aged 15} & 1st & .096^{**} & .092^{**} & .086^{*} \\ 
outcome: \text{Probability of holding college degree}	 & 2nd & 2.012^{**} & 1.267 & -.837 \\ \hline

\end{array}$
\label{results}
\text{$*$ for significant at $10\%$, $**$ for significance at $5\%$ and $***$ for significance at $1\%$.}
\end{table}

\begin{figure}[H]
\includegraphics[width=\textwidth] {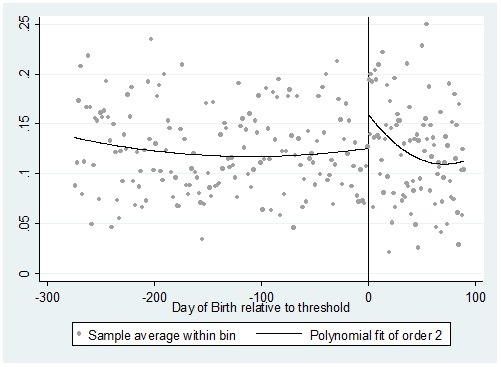}
\caption{Probability of holding a college degree and date of birth for people between 16 and 34 who lived in a urban city when aged 15.}
\label{escolaridade_2_descontinuidade urbano 15 faculdade}
\end{figure}

However, surprisingly, we do find this discontinuity in our third sample, as showed in figure~\ref{escolaridade_2_descontinuidade urbano 15 faculdade}, and it is statistically significant. When used as an instrument for occupied population, it has an effect of 0.096 on years of study, and, in the second stage, holding a college degree is shown to have an significant impact of 201\% on labor income, but not on labor supply neither probability of being occupied.

\section{Robustness checks}\label{robustez}

There are three assumptions for any Regression Discontinuity Design to be considered robustly giving a estimate of causal effects. First, individuals need not to be able to control precisely the assignment variable - the Day of Birth, in this case. Besides, all other factors must be “continuous” with respect to the independent variable - that is, years of study or probability of holding a college degree. Finally, the parametric functional form need to be correctly fit into the function of the dependent variable. In this paper, we address robustness checks for the first two assumptions.

\cite{betran2016increasing} shows that, in a sample of 150 countries in 2014, Brazil has a cesarean section rate of 55.6\%, the highest one in Latin America and the Caribbean, which has the highest rates in the world. Qualitative data suggests that one of the reasons for some Brazilian mothers rather have this procedure is due to the possibility of time of birth \citep{kasai2010women}. If mothers are able to manipulate when their children are going to be born by scheduling a cesarean section, and the possibility of this procedure and women preferences for day of birth of their children is not a smooth function around the cutoff, our estimates may be biased because of mother's choices. So, we'll present a histogram of day of birth relative to March 31 and apply a McCrary Density Test \citep{mccrary2008manipulation} to check for any evidence of manipulation of the treatment variable.

Besides, following the methodology of \cite{ferreira2003mobilidade}, we use categorical data on education level of parents of the population when they were 15 years old, from the supplement, to construct a variable for parental schooling. Then, we test for any discontinuity in these variables around the cutoff, checking if the increase in the years of study of children just before March 31 may be due to more educated parents on that side of the threshold.  

In order to give additional support to our hypothesis that being born just before the primary school entry threshold implies an exogenous variation in schooling, we will also make use of placebo variables, specifically age and sex, so we can if there is any inconsistency in the causality link. Finally, check for other possible primary school entry dates as instruments for the population that used to live in urban areas when 15, aged between 16 and 34.

\begin{figure}[H]
\includegraphics[width=\textwidth] {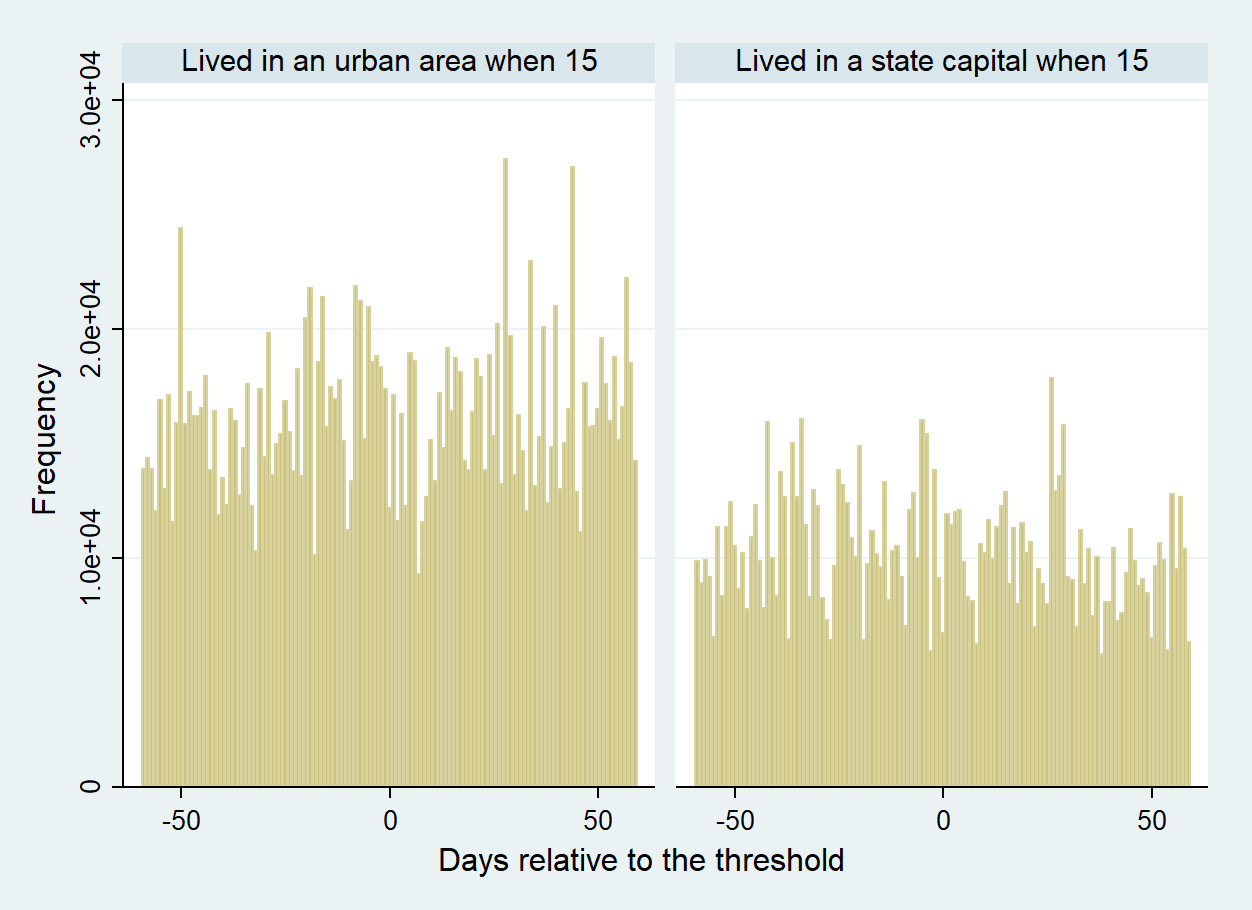}
\caption{Date of birth frequency for people between 16 and 34 who lived in a urban city when aged 15.}
\label{escolaridade_2_descontinuidade frequencia}
\end{figure}

\begin{figure}[H]
\includegraphics[width=\textwidth]{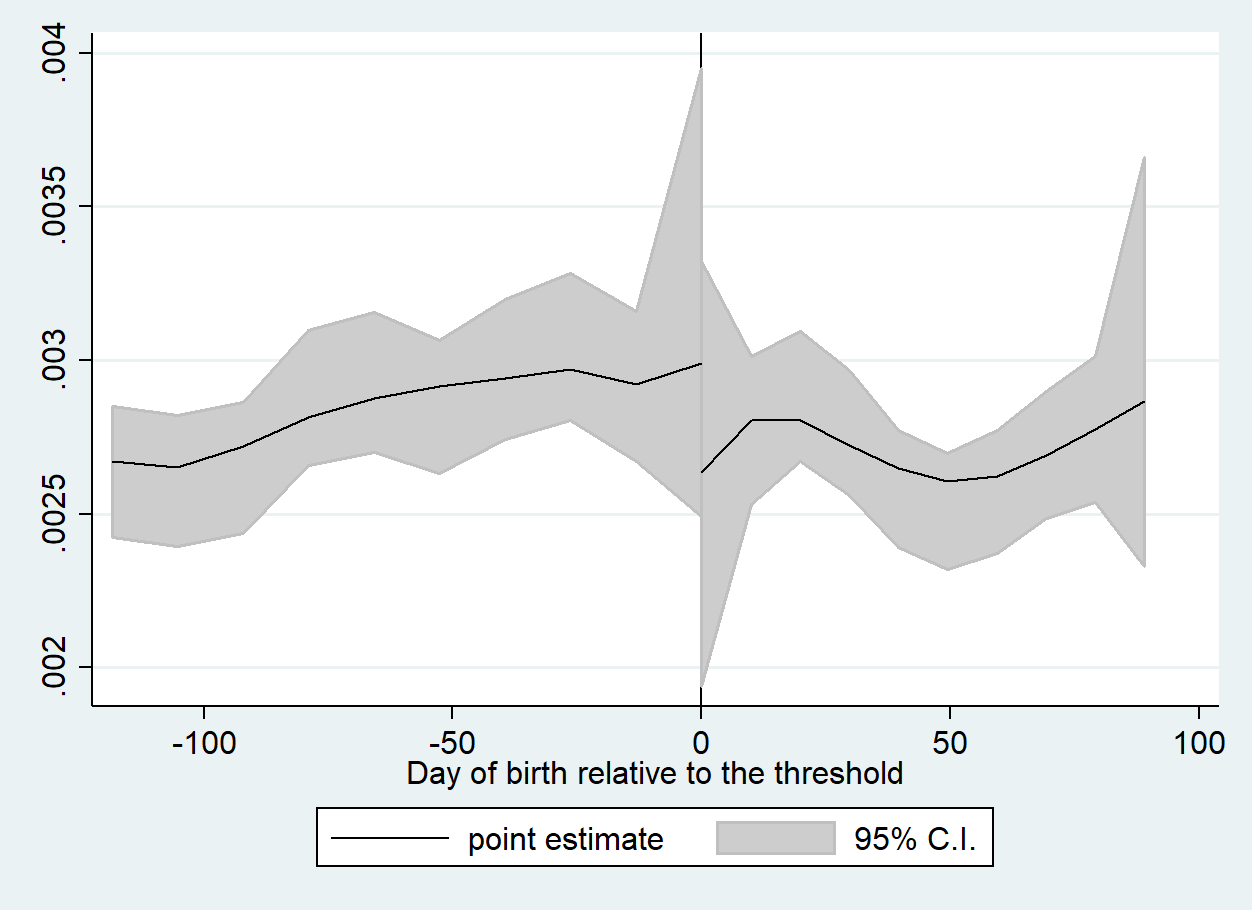}
\caption{Density of birth dates for people who lived in a state capital when aged 15.}
\label{densidade_capital}
\end{figure}

\begin{figure}[H]
\includegraphics[width=\textwidth]{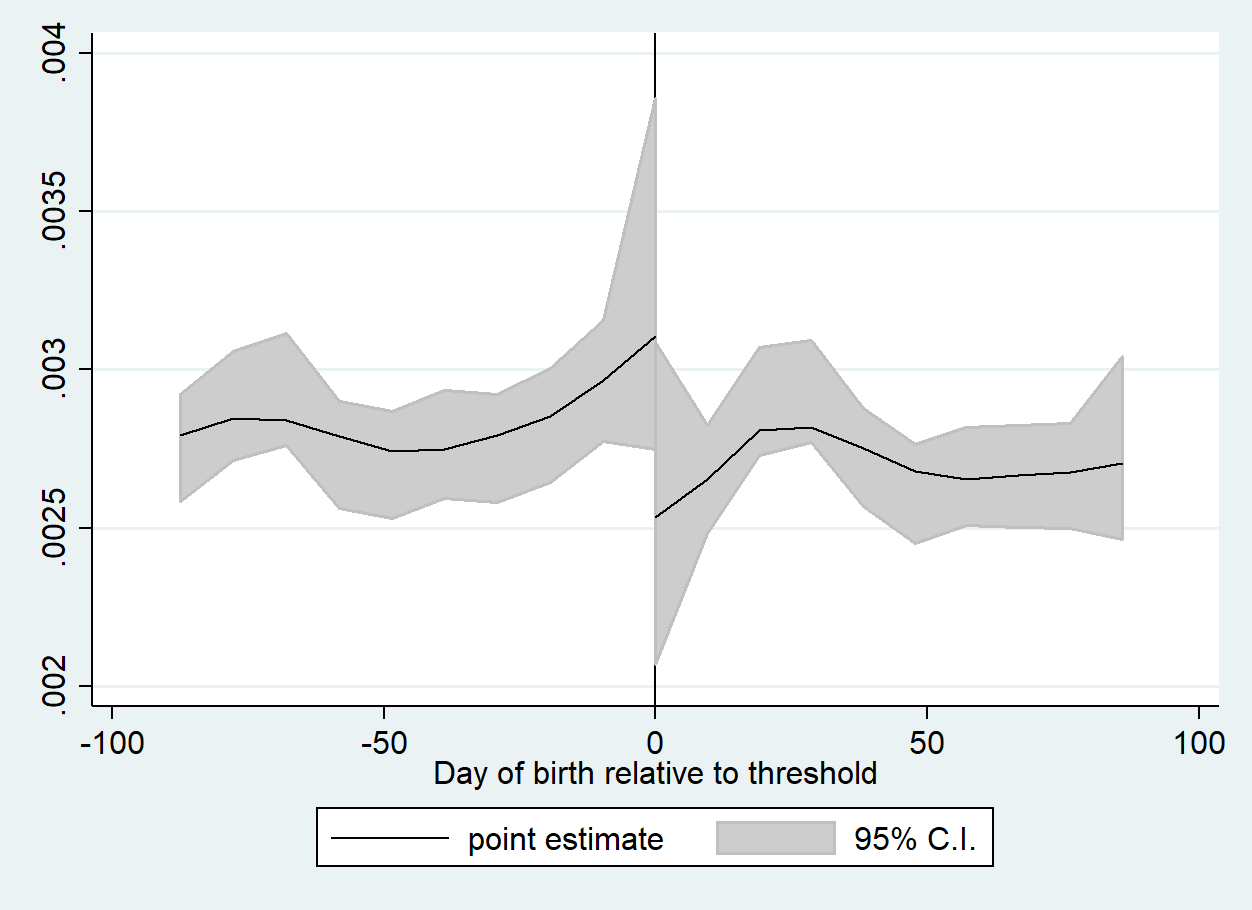}
\caption{Density of birth dates for people who lived in an urban area when aged 15.}
\label{densidade_urbanos}
\end{figure}

First, figure~\ref{escolaridade_2_descontinuidade frequencia} shows a relative minimum of birthdays at the 0 cutoff for all population under 35 who lived in state capitals or urban areas when aged 15. However, applying a McCrary Density Test, we do not find a estimated discontinuity in the density function of the running variable at the cutoff statistically significant at 5\%, as shown in figures~\ref{densidade_capital} and~\ref{densidade_urbanos}. Table~\ref{datas_corte_estaduais_e_fakes} shows that, regarding parental educational level, there is no statistically significant difference between the population born just before and just after the most common primary school entry date. This results suggests that, even if there was manipulation on date of birth, parents more or less educated do not differ when choosing to and how to manipulate it around March 31. 

\begin{table}[H]
\caption{Placebos, fake thresholds and RDD results}
 $\begin{array}{c c c } 
 \hline
 \text{Date} & \text{RDD Estimate} & \text{p-Value}  \\ [0.5ex] 
 \hline\hline\addlinespace
 \text{Threshold} & .987 & .001  \\ [0.5ex]
 \addlinespace
  \text{Placebo Threshold 2 - February 28 } & -.223  & .595 \\ [0.5ex]
 \addlinespace
 \text{Placebo Threshold 1 - April 30} & 1.462 & .000  \\ [0.5ex]
 \addlinespace
 \text{Placebo Threshold 3 - May 30} & .224 & .359  \\ [0.5ex]
 \addlinespace
 \text{Placebo Threshold 4 - June 30} & .066 & .817 \\ [0.5ex]
 \addlinespace
 \text{Mother's Education} & .276 & .635 \\ [0.5ex]
\addlinespace
 \text{Father's Education} & .882 & .148 \\ [0.5ex]
\addlinespace
\text{Female} & .034 & .583 \\ [0.5ex]
\addlinespace
 \text{Age} & .860 &  .134 \\ [0.5ex]
\addlinespace
\hline
 \end{array}$
\label{datas_corte_estaduais_e_fakes}
\\[1.5pt] 
\end{table}

Our placebo tests for age and sex show no discontinuity before or after the threshold. However, testing for discontinuities with "sharp" RDDs in other dates, we do find a sharp rise in Years of Study to Brazilian born just before April 30. That is maybe due for this date to be another usual Primary School entry date rule, as it is legislated in Mato Grosso state and Atibaia city. For every other possible primary school entry date rule before July, we do not find any discontinuity.

\section{Conclusion}\label{conclusão}

In this paper, we have shown that, using socio-occupational mobility data from a supplement of the 2014 Edition of Brazilian National Sample Household Survey (PNAD), it is possible to identify an credibly exogenous increase in educational level for the population between 16 and 34 years old who used to live in state capitals or urban areas when 15 years old, due to an arbitrary primary school entry date policy. Also, we have argued that, when considering the larger sample, this may be used as an instrument for marginal increase in education, having an impact on labor income of 25.8\%. 

\cite{10.1257/aer.101.1.158} argue that discontinuities in education around school entry date at adulthood happen naturally for those who are still in school or dropouts, who have fewer years of schooling if they got enrolled in primary school one year later. But, beyond these natural channels, we have shown that being born just before the threshold also has an impact on the probability of holding a college degree in adulthood. Using this discontinuity as instrument, we find a tertiary education return of 201\% on labor income. 

The estimates found in this paper are unusually higher than this literature in developing economies. That might happen because of the possible negative selection of our treated group, since their parents have complied with the primary school entry policy rules, what is expected to be associated with lower socioeconomic status. 

We have also argued that, despite Brazil being one of the countries with the highest cesarean section rate in the world, there is little evidence of manipulation on the day of birth in our samples. We also see no statistically significant difference on parent's education just before and after the threshold, suggestion that richer and poorer parents alike aren't manipulating birth dates - at least not with entry dates in mind. 

Summarizing, our results sustain two conclusions: first, enrolling a child in primary school one year earlier has an significant long term impact on educational outcomes. The most significant one being the increase in probability of having a college degree, that rises around 9.6\% due to being born just before March 31. Secondly, education still has significantly large impact on labor income in Brazil, around 25\% for one more year of study and 200\% for having tertiary education. This result is likely representative of other lower-middle income economies. Then, that educational performance might explain a high share of income inequality, as sustained by previous literature on this topic.

\clearpage
\bibliographystyle{apa}
\bibliography{bib_schooling}

\end{document}